\documentclass[pra, twocolumn, showpacs, superscriptaddress]{revtex4}

\usepackage{amssymb}
\usepackage[dvips]{graphicx}
\usepackage[usenames,dvipsnames]{color}
\usepackage{amsmath}

\newcommand{\bra}[1]    {\langle #1|}
\newcommand{\ket}[1]    {| #1 \rangle}

\begin{document}

\title{Long Distance Entanglement Generation in
 2D Networks}

\author{S.~Broadfoot}
\affiliation{Clarendon Laboratory, University of Oxford, Parks Road, Oxford OX1 3PU, United Kingdom}
\author{U.~Dorner}
\affiliation{Centre for Quantum Technologies, National University of Singapore, 117543, Singapore}
\affiliation{Clarendon Laboratory, University of Oxford, Parks Road, Oxford OX1 3PU, United Kingdom}
\author{D.~Jaksch}
\affiliation{Clarendon Laboratory, University of Oxford, Parks Road, Oxford OX1 3PU, United Kingdom}
\affiliation{Centre for Quantum Technologies, National University of Singapore, 117543, Singapore}

\date{\today}

\begin{abstract}
  We consider 2D networks composed of nodes initially linked by
  two-qubit mixed states. In these networks we develop a global error
  correction scheme that can generate distance-independent
  entanglement from arbitrary network geometries using rank two
  states. By using this method and combining it with the concept of
  percolation we also show that the generation of long distance
  entanglement is possible with rank three states.
 Entanglement percolation and global error correction
    have different advantages depending on the given 
    situation. To reveal the trade-off between them we consider their
    application on networks containing pure states. In doing so we
  find a range of pure-state schemes, each of which
    has applications in particular circumstances: For instance, we
  can identify a protocol for creating perfect entanglement between
  two distant nodes. However, this protocol can not generate a singlet
  between any two nodes. On the other hand,
  we can also construct schemes for creating entanglement between
  any nodes, but the corresponding
  entanglement fidelity is lower.
\end{abstract}

\maketitle

\section{Introduction}  \label{sec1}

Quantum repeaters have shown great promise for
  creating entangled states over long distances, which is essential
  for many quantum processing
  tasks~\cite{Briegel98,Duan01,Duer99,Hartmann07,Dorner08}. The
  repeaters consist of a 1D network of nodes linked by entangled
  states. By using local operations and classical communication (LOCC)
  an entangled state is generated between distant nodes, with the
  required number of initial entangled states between each node
  scaling logarithmically with the states separation
  distance~\cite{Jiang09}. Once created the entangled state can be
  used for a range of tasks including quantum
  cryptography~\cite{Bennett84,Ekert91}, distributed quantum
  computing~\cite{Cirac99} and quantum
  teleportation~\cite{PhysRevLett.70.1895}. The goal of these protocols is to generate a highly entangled state between two distant nodes and typically one uses the states fidelity, i.e. overlap with a singlet, as a measure of success in achieving this.
  
  However, the one dimensional nature of quantum
  repeaters means that their application in higher dimensional
  networks will not take advantage of the higher connectivity. By
  creating protocols for these higher dimensional networks the required amount of initial entanglement between the nodes can be reduced. This was first discussed for pure state networks and involved the use of `entanglement
  percolation'~\cite{Acin07,Perseguers08,Lapeyre09,Cuquet09,Perseguers09b,Eisert07}.
  The procedure uses the `procrustean method' to transform each pure
  state into a maximally entangled singlet with a finite
  probability~\cite{Bennett96}. When this probability exceeds a
  geometry dependent threshold an infinite cluster is formed of linked
  nodes. Any two nodes in the cluster can then be transformed into a
  singlet using the process of entanglement
  swapping~\cite{Bose99,Perseguers08}. A singlet is generated between
  nodes with a probability that is independent of their separation. In
  particular, the required number of entangled states between each node does not
  change with the separation, in contrast to quantum repeaters.

Unfortunately, in reality every state will experience noise that
causes it to lose its purity and become mixed. Therefore extending these
ideas to mixed states is of vital importance for applications. In
previous work we have extended entanglement percolation to networks
that initially consist of a particular type of mixed
state~\cite{Broadfoot09,Broadfoot10}, that can occur when channels
are subject to amplitude damping, and we have shown that these are the only
two-qubit states that allow perfect singlets to be created.

Other schemes have also been shown to generate highly entangled states from a regular 2D square network made of `binary
states'~\cite{Duer99,Perseguers08b}. Here, if the amount of entanglement
in the binary states exceeds a threshold, the network is transformed
into an entangled state that can stretch over an arbitrary distance,
while maintaining a non-maximal but constant entanglement. Our
previous protocol~\cite{Broadfoot09}, as well as the scheme described
in Ref.~\cite{Perseguers08b}, requires states of rank two, or less, for
long distance entanglement generation using constant resources. So far
no protocol has been able to transform a 2D network of full rank
states into a highly entangled two qubit state with no dependence on
the qubit separation, although this is possible in infinite 3D
networks~\cite{Perseguers09c}.

In this paper we extend on these ideas to obtain a `global error
correction' procedure that can be applied to quantum networks that
have arbitrary geometry and are composed of binary states. By
combining it with quantum state percolation we will show that
entanglement can be efficiently generated over long distances by using
a constant number of \emph{rank three} states between each node. The
final fidelity is independent of the distance. Pure
  states can also be used by the global error correction and
  entanglement percolation protocols. These can again be combined to
  create a range of altered schemes
  that allow for a continuous transition from a pure state global
  error correction method to entanglement percolation. We find that
  global error correction allows any two chosen nodes in the network to be linked by an entangled
  state, but this is at the expense of the state's fidelity.
  Entanglement percolation only allows two nodes from a known subset to be to be connected,
  but this is via a perfect singlet. If the criteria for entanglement
  percolation are not satisfied we have to use a specific combined scheme to create
  states of maximal fidelity and the choice of this scheme is dependent on the initial states in the
  network.

The paper is structured as follows. In Sec.~\ref{sec2} we define the
network and develop a generalized form of global error correction.
This scheme works on binary states and in Sec.~\ref{sec3} we give
a way of generating these binary states from rank three states. These
capabilities are then combined to enable long distance entanglement
distribution with rank three states. Requirements for the scheme are
calculated and compared with numerical results. In Sec.~\ref{sec4} the
schemes will be altered to work on pure states and this allows
comparisons to be made between the protocols. Finally we conclude in
Sec.~\ref{sec5}.

\begin{figure}[]
  \centering\includegraphics[width=8cm]{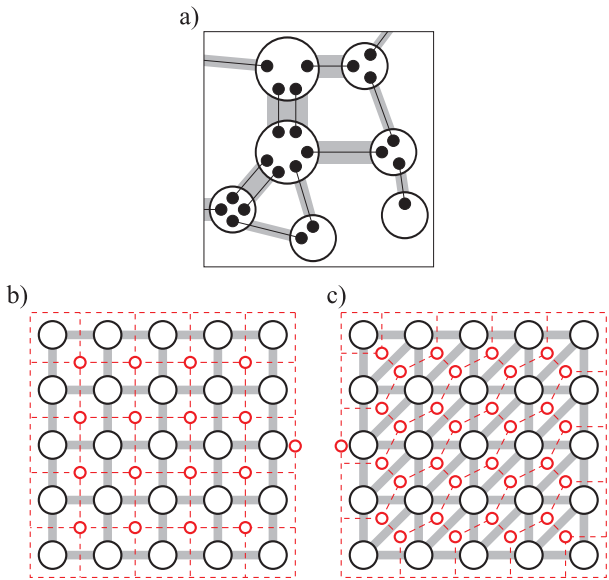}
\caption{%
(color online) a) A quantum network is shown composed of nodes (large circles) that contain qubits (black circles). Each qubit is initially entangled to another (thin black line) and these two qubit states form the edges that are contained within each bond (thick grey lines). b) A 5x5 node square network with its dual network. The individual edges and qubits are not shown. However, the dual nodes are presented as red, small circles connected by dashed lines, including the network exterior (dashed square with small circle). c) A 5x5 node triangular network with its honeycomb dual network.}
\label{fig1}
\end{figure}

\section{Global Error Correction \label{sec2}}

   A quantum network is constructed by sharing two-qubit entangled states between locations that we call `nodes'. All quantum operations must be local to these nodes but classical communication is allowed between them. The shared two-qubit states are referred to as `edges' and all of the edges between two nodes constitute a `bond' (see Fig.~\ref{fig1}). Ideally each edge will be a maximally entangled singlet, however due to decoherence this situation is unrealistic and we must have a method to cope with errors. In this section we will devise a global error correction method that performs local measurements and, by combining the results, information is obtained on the error locations. This information does not allow us to exactly locate the errors but, as we will see, does allow the errors to be subdued so entanglement can be created between distant nodes. Note that throughout we assume the measurements and operations act perfectly.

\begin{figure}[]
  \centering\includegraphics[width=6cm]{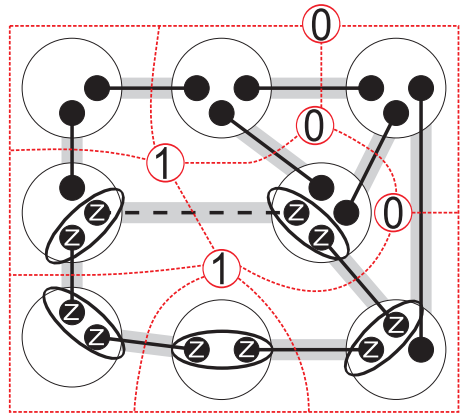}
\caption{%
(color online) One possible pure state network $\rho_{\left\{a_i\right\},\left\{b_i\right\}}$ from the ensemble given by Eq.~\ref{eq1}. Each node (white circles) contains qubits (black dots) that are part of two qubit entangled states (thin lines) and edges are located in bonds (thick grey lines). For $a=0$ the edge line is solid and if $a=1$ it is dashed. $Z \otimes Z$ measurement operations are carried out within each node (black ovals) around one plaquette. By doing this a value of $L_{loop}$ can be associated with each plaquette. The dual network (red dashed lines) is shown with these values contained within each dual node (red circle). The dual vertex representing the exterior is represented by a dashed box.}
\label{fig2}
\end{figure}

   We will consider the entire network as a mixture of pure states and analyze the protocol acting on each of these using the stabilizer formalism~\cite{NielsenChuang}. From these results we can then determine the mixed state produced. To begin we construct the network's initial state using the states of individual edges. Each of these initial two-qubit edge states can be transformed into a probabilistic mixture of Bell states~\cite{Bennett96a}, via LOCC. This allows us to assume that each edge is in one of the four Bell states given by $\ket{\psi_{ab}}=(X^a Z^b\otimes I) (\ket{00}+\ket{11})/\sqrt{2}$, where $a,b\in\left\{0,1\right\}$. A stabilizer generated by $\left\{(-1)^a Z\otimes Z,(-1)^b X\otimes X\right\}$ can then describe each Bell state and we label the probability of that state as $p_{ab}$. For a network that contains $N_E$ of these identical edges, and $N_Q=2 N_E$ qubits, the whole state will have the form
\begin{align}
          \rho_I &= \sum_{\left\{b_i\right\}} \sum_{\left\{a_i\right\}} \prod_{i=1}^{N_E} p_{a_i b_i} \ket{\psi_{a_i b_i}}\bra{\psi_{a_i b_i}}\nonumber \\
          &= \sum_{\left\{b_i\right\}} \sum_{\left\{a_i\right\}} P(\left\{a_i\right\},\left\{b_i\right\})\rho_{\left\{a_i\right\},\left\{b_i\right\}},
\label{eq1}
\end{align}
where each edge, $i$, has parameters $a_i\in\left\{0,1\right\}$ and $b_i\in\left\{0,1\right\}$. The summations are over all values that these can take and give a mixture of pure states
\begin{equation}
\rho_{\left\{a_i\right\},\left\{b_i\right\}}=\prod_{i=1}^{N_E} \ket{\psi_{a_i b_i}}\bra{\psi_{a_i b_i}},
\label{eq2}
\end{equation}
that occur with probability
\begin{equation}
P(\left\{a_i\right\},(\left\{b_i\right\})=\prod_{i=1}^{N_E} p_{a_i b_i}.
\label{eq3}
\end{equation}
The error model we consider has independent bit-flip and phase-flip errors. For this case the pure states that contribute to the mixture can be thought of as networks of error free singlet states, $\ket{\psi_{00}}$, on which errors randomly occur. A bit-flip error that causes $a\rightarrow 1$ occurs with probability $p_x$. Similarly, phase-flip errors cause $b\rightarrow 1$ and have a probability of $p_z$. This situation is equivalent to having edges with the values $p_{00}=(1-p_x)(1-p_z),\,p_{01}=(1-p_x)p_z,\,p_{10}=(1-p_z)p_x,$ and $p_{11}=p_x p_z$ and it simplifies the probability of each state $\rho_{\left\{a_i\right\},\left\{b_i\right\}}$ to become
\begin{align}
\nonumber \\
P(\left\{a_i\right\},(\left\{b_i\right\})&=p_x^{N_X}(1-p_x)^{N_E-N_X}\times  \nonumber \\
& p_z^{N_Z}(1-p_z)^{N_E-N_Z},
\label{eq4}
\end{align}
with $N_X$ and $N_Z$ giving the number of bit-flip and phase-flip errors on the state, respectively,
\begin{align}
\nonumber \\
N_X &= \sum_{i=1}^{N_E} a_i, \nonumber \\
N_Z &= \sum_{i=1}^{N_E} b_i.
\label{eq5}
\end{align}
Now we apply the global error correction procedure to one of these
pure states. Each of which can be described using a stabilizer that is
formed by the union of Bell state stabilizers. The procedure extracts
information by performing $Z\otimes Z$ measurements locally on nodes
along a closed path of edges, as illustrated in Fig.~\ref{fig2}. Such
paths are referred to as `loops' and contain edges from the set
$loop=\left\{l_1,l_2,....l_{N_L}\right\}$, where the $l_1,l_2...$ give
the index of each of the $N_L$ edges forming the loop. These measurements take $-1$ and $1$ as possible values. The product of
the measurement results around a loop is labelled $(-1)^{L_{loop}}$. Here $L_{loop}$ gives the bit-flip error parity around the loop, $L_{loop}= \left(\sum_{i\in loop} a_{i}\right) mod \,2$. Each value of $L_{loop}$ gives us information on the errors that we
can then use. As the measurements are made the stabilizer follows the
rules given in Ref.~\cite{NielsenChuang}. After each measurement the
stabilizer generator is manipulated so that at most one element
anti-commutes with the measurement operator and the measurement
operator, multiplied by the outcome, is then substituted in place of
the anti-commuting element. If the operator does not anti-commute with
any elements in the stabilizer then nothing is changed.

\begin{figure}[]
  \centering\includegraphics[width=5cm]{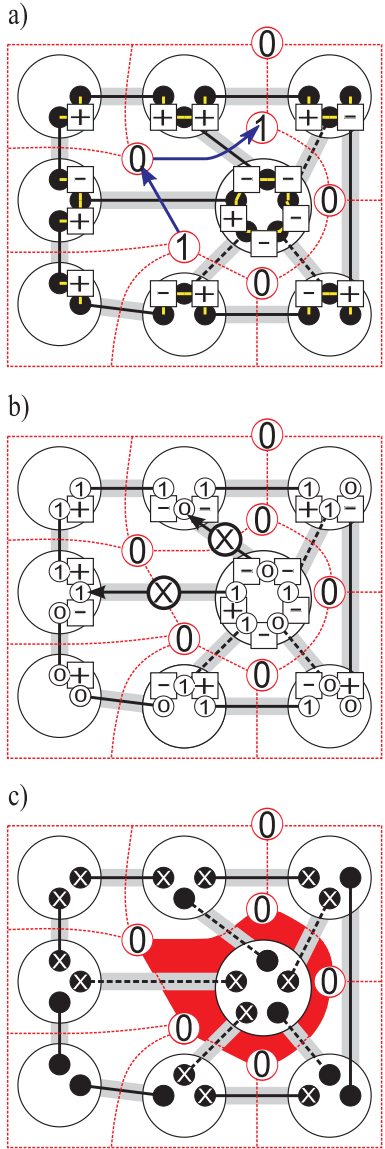}
\caption{%
(color online) a) A single edged network is shown where three edge errors have occurred (dashed lines). The local $Z \otimes Z$ measurement outcomes are displayed inside boxes, with `$\pm$' referring to a $(-1)^{m_j}=\pm1$ eigenstate. Around each plaquette the values of $L_{loop}$ are then calculated and shown on the dual network nodes (red outlined circles linked by dotted lines). Values of $1$ are defects and are matched together along a path (blue arrows). b) Edge errors are assumed to lie across the matching path and X gates are applied to one qubit within each of these edges (black arrows). This `flips' ($+\leftrightarrow -$) some of the local measurement outcomes. Each qubit can then be put into two groups (labelled `0' and `1') using the measurement outcomes (`-' means the groups need to be different for the two qubits) and qubits that are in a network edge get put in the same group. To assign the values we pick an initial qubit and value at random and then assign values in all remaining qubits. c) X gates are applied to one group (nodes containing white X's) and for each pure ensemble state a GHZ state is created with patches (shaded red area) of bit-flip errors. These occur where the actual edge error path differed from the assumed path.
}
\label{fig3}
\end{figure}

We obtain the measurement outcomes, $(-1)^{m_j}$ ($m_j\in\left\{0,1\right\}$), for the operator $Z \otimes Z$ between every pairing of qubits in each node. By making these measurements we can calculate $L_{loop}=  \left(\sum_{i\in loop} m_{i}\right) mod \,2$ for all of the smallest possible loops in the network, which we refer to as plaquettes, and can then deduce $L_{loop}$ for any loop. We do not need to perform the actual measurement between every qubit pair in a node because some of the $Z \otimes Z$ operators will be products of previously measured $Z \otimes Z$ operators. From the state $\rho_{\left\{a_i\right\},\left\{ b_i\right\}}$ these measurements result in a state given by a stabilizer that has generators $(-1)^{N_Z}(X^{\otimes N_Q})$, $(-1)^{a_i}(Z\otimes Z$) for every edge, $i$, and operators $(-1)^{m_j} (Z\otimes Z)$ for every measurement $j$.

\begin{figure}[]
  \centering\includegraphics[width=8cm]{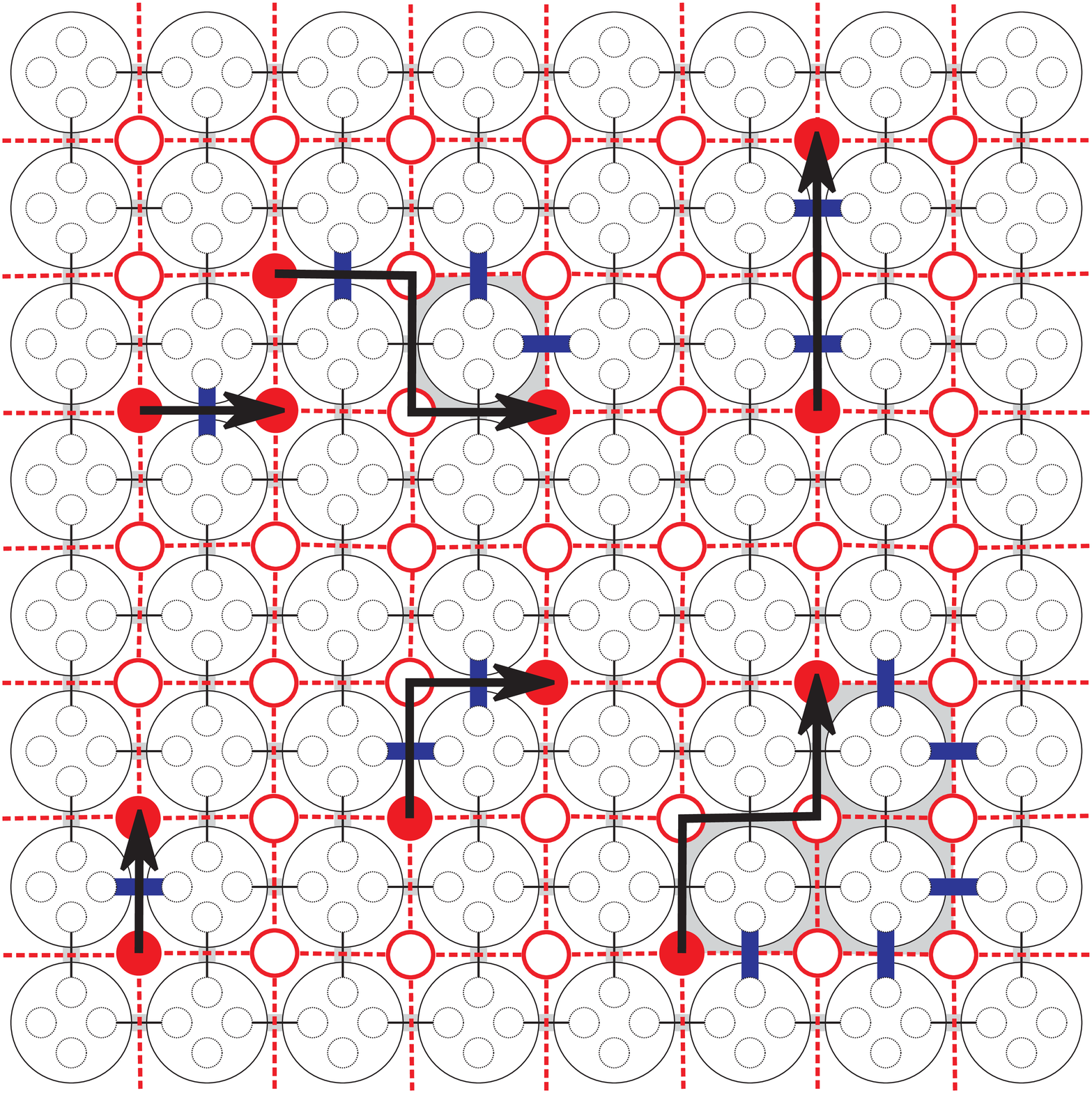}
\caption{%
  (color online) Regular square network formed by single edged bonds.
  The dual lattice is shown in red/dashed and includes the error
  syndromes on the dual vertices (red, small circles; $L_{loop}=1$-filled, $L_{loop}=0$-outlined). The edge errors causing these plaquette errors are shown
  by thick blue lines. The error syndromes are matched together (black
  arrows) and this can be used to give the most likely edge error
  locations for the recorded measurement syndrome. The grey areas
  between the assumed error locations and actual edge errors
  identifies groups of nodes. For this particular pure state in the
  ensemble we will obtain a GHZ state with X errors present on the
  qubits within these nodes. In practise we do not know what the initial state
  was and hence do not know the location and size of the grey patches, but for the regions to become larger more errors are required and hence are less likely.}
\label{fig4}
\end{figure}

All of the plaquettes in the network can then be assigned their values
of $L_{loop}$. Values of one can be considered as defects on nodes in the dual network (see Fig.~\ref{fig2}). These defects occur in pairs. A chain of errors, $a_i=1$, on the quantum network's edges separate the dual node defects in each pair (see Fig.\ref{fig3}a). These error affected edges must cross a path of dual network edges that connect the two defects. Hence, we can obtain information on the location of bit-flip errors by
pairing the defects, in a matching problem. We assume that the length
of the paths is minimal, since longer paths require more errors and
this is unlikely when $p_x$ is small. This gives us a minimum weight
matching problem to be solved \cite{Edmonds65} in an analogous way to
the correction methods used in surface codes
\cite{Kitaev97,Kitaev03,Dennis02}. By doing so we have found the most
likely defect pairs to be linked by a path of errors. The actual path
is assumed to be the one of shortest distance, i.e. smallest number of
edges, linking the defects on the dual network (see Fig.\ref{fig3}a).
Sometimes more than one such path exists and we would wish to minimize
the number of nodes between the true path and the assumed path. This
can be easily done on regular networks, by using paths that
approximately follow straight lines, but for simplicity in the general
case it is sufficient to choose one of the shortest paths at random.
One qubit in each network edge that lies along the assumed paths is
then acted on with a X gate. We can actually avoid performing these X gates until the very end when we may have fewer qubits to apply gates on. Here we include the gates at this stage for clarity. When these gates are applied to a qubit we also change the value of $m_j$ to $1-m_j$ for each measurement pairing that includes those qubits (see Fig.\ref{fig3}b). This gives $m_j=1$ for an even number of times around every loop and by acting around every loop
with $X$ gates we can force $m_j=0$ for every local qubit pairing. A
procedure to accomplish this task is given in Fig.\ref{fig3}(b-c).
Once this is done the resulting stabilizer will have generators
$(-1)^{N_Z} X^{\otimes N_E}$, $(-1)^{a_i'}Z\otimes Z$ for every edge,
$i$, and $Z \otimes Z$ from every local pairing of qubits. The $a_i$
values have been changed to $a_i'$ after performing the $X$ gates.
These $a_i'$ values are unknown, however all of the dual edges that
cross edges with $a_i'=1$ must form loops on the dual network. This
stabilizer describes a pure GHZ state that has had $N_Z$ Z operators
acting on it and X operators acting on all the qubits within some
nodes. These affected nodes are within loops on the
dual network and appear as patches (see Fig.~\ref{fig4}). The X gates
that remain when the procedure is run on an initial configuration
$\left\{a_i\right\}$ are described by an operator
$X_P(\left\{a_i\right\})$, that applies X gates on the qubits that lie
within patches. The whole procedure will result in a mixture of these
pure states,
\begin{align}
\rho_F = \sum_{\left\{b_i\right\},\left\{a_i\right\}} & p_z^{N_Z}(1-p_z)^{N_E-N_Z}P(\left\{a_i\right\} | \left\{m_j\right\}) \times  \nonumber \\
& X_P(\left\{a_i\right\}) Z^{N_Z}\rho_{GHZ} Z^{N_Z} X_P(\left\{a_i\right\}),
\label{eq6}
\end{align}
where $Z$ acts on an arbitrary qubit and $P(\left\{a_i\right\}| \left\{m_j\right\})$ is the probability of the initial error configuration being $\left\{a_i\right\}$ given the results of the measurements.

Larger patches require more errors and are less likely to occur.
Hence, for low values of $p_x$ the most probable case is that 
finite sized patches, containing a small number of nodes, occur randomly. We quantify
the probability of a random node existing in a patch as $P_X$, which
has the same value throughout the network. For larger values of $p_x$
the likely number of errors increases so that it becomes impossible to
match the defects up correctly and accurately guess the paths. Within infinite networks there exists a critical value of $p_x$ below
which patches almost certainly exist and long distance entanglement 
can be generated. Above the critical value this is not
possible and $P_X=1/2$ for the most likely pure states.

\begin{figure}[]
  \centering\includegraphics[width=6cm]{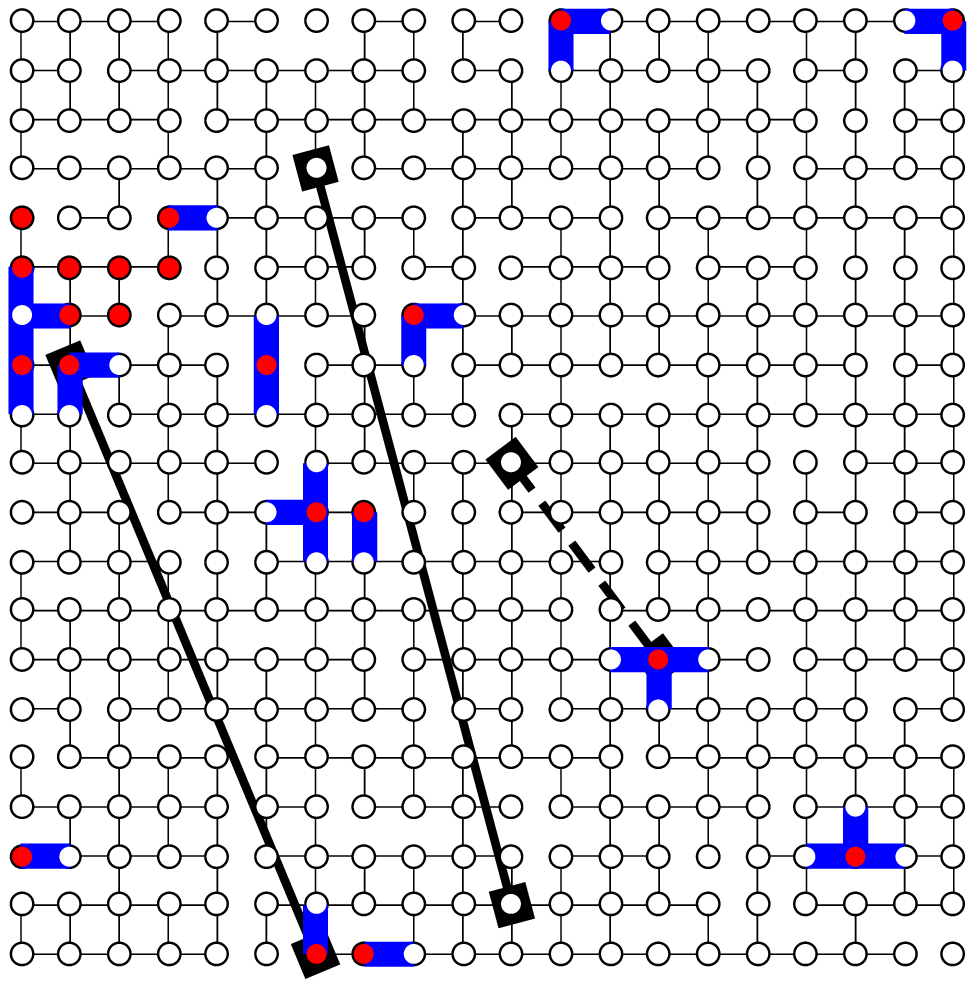}
\caption{%
(color online) Here a 20x20 network has had its bonds removed with a probability of $0.05$ and then X errors introduced with a probability of $0.05$. We apply our global error correction procedure and eventually this generates a GHZ state with patches of nodes that exhibit unwanted, extra bit-flips (red, filled circles). If we try to link a node from within a patch to outside of them (dashed line) the final state two-qubit state exhibits a bit-flip error. Examples are also given of qubit pairings that would not exhibit errors (bold black lines).
   }
\label{fig5}
\end{figure}

When we are highly likely to form patches the resulting mixed state can be cut back to an impure GHZ state, with any number of qubits up to $N_Q$, by removing qubits with X measurements and, depending on their outcomes, applying a correcting Z operation on one of the qubits to be kept. The final fidelity, i.e. overlap with a pure GHZ state, is dependent on $p_x$ and decreases for more qubits in the resulting state. It is however still independent of the distance between the qubits. In particular, we can remove all of the qubits except two leaving a two-qubit entangled state spread between two nodes. For each pure state contributing to the mixture in Eq.~(\ref{eq6}) after removing the qubits we obtain an ideal singlet when the remaining qubits are either both in an error affected patch or outside of them (see Fig.~\ref{fig5}). The pure state $\rho_{\left\{a_i\right\},\left\{b_i\right\}}$ is transformed to the state $\rho_c(\left\{b_i\right\},\left\{a_i\right\})$, that has the form
\begin{align}
          \rho_c= &(P_X^2+(1-P_X)^2)\frac{(1+(1-2p_z)^{N_E})}{2} \ket{\psi_{00}}\bra{\psi_{00}} \nonumber \\
          &+ 2 P_X (1-P_X)\frac{(1+(1-2p_z)^{N_E})}{2}\ket{\psi_{10}}\bra{\psi_{10}}  \nonumber \\
          &+ (P_X^2+(1-P_X)^2)\frac{(1-(1-2p_z)^{N_E})}{2} \ket{\psi_{01}}\bra{\psi_{01}} \nonumber \\
          &+ 2 P_X (1-P_X)\frac{(1-(1-2p_z)^{N_E})}{2}\ket{\psi_{11}}\bra{\psi_{11}},
\label{eq7}
\end{align}
and the final mixed state is given by the weighted sum of these,
\begin{align}
\rho_{f} = \sum_{\left\{b_i\right\},\left\{a_i\right\}} & p_z^{N_Z}(1-p_z)^{N_E-N_Z}P(\left\{a_i\right\} | \left\{m_j\right\}) \times  \nonumber \\
& \rho_c(\left\{b_i\right\},\left\{a_i\right\}).
\label{eq8}
\end{align}

\begin{figure}[]
  \centering\includegraphics[width=8cm]{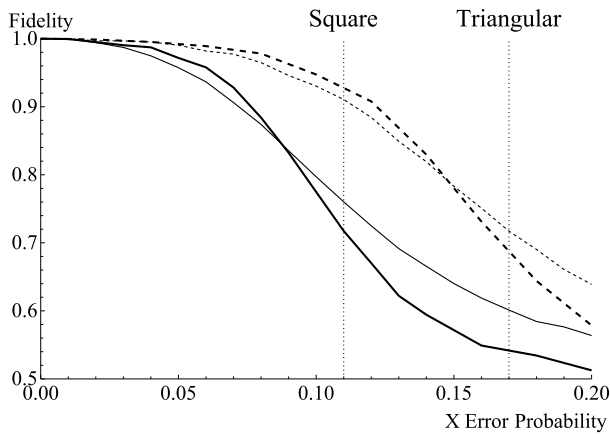}
\caption{%
The fidelities are plotted from simulations of global error correction when run on complete square and triangular networks to produce a two qubit entangled state between two randomly chosen qubits. In the square case (solid lines) we see a lower threshold compared to the triangular network (dashed lines). We also included the approximate threshold values (dotted vertical lines) that are calculated in Sec.~\ref{sec3}. We used networks of 10x10 nodes and 20x20 nodes, with the larger networks presented by thicker lines, and can see that the threshold becomes more pronounced when the network gets larger.
   }
\label{fig6}
\end{figure}

Since $P_X$ is independent of the state separation we see that as long
as there are no phase errors present, i.e. $p_z=0$, the state produced
has a fidelity that is independent of the node separation. These states that do not experience phase errors are called
binary states~\cite{Duer99}. Our global error correction procedure is
a generalization of the one given in \cite{Perseguers08b} but does not
involve teleporting qubits and can be easily applied to any network
geometry. Later we will see that we can randomly transform states into binary states, obtaining them on random bonds but also destroying
  all of the entanglement in the remaining bonds. Our scheme can still be
applied in these situations and in Fig.~\ref{fig5} it has been applied on
a contributing term in the largest cluster of a square network that
was missing such bonds. We will discuss this in more detail in the next
section (see Sec.~\ref{sec3}). Furthermore, in regular networks with all bonds intact we performed simulations to reveal the
critical threshold of $p_x$ below which long distance entanglement distribution is always possible. This was done for both square and triangular
networks (see Fig.~\ref{fig1}(b-c)). In both cases we see the final
states fidelity decrease suddenly after critical $p_x$ values (see
Fig.~\ref{fig6}). These transitions become more prominent for larger
networks and later we will attach values to the thresholds.

If we have both bit-flip and phase errors present the initial edges will be rank four Werner states~\cite{Werner89}. To apply the procedure in this case we can use error correction codes to subdue the phase-flip errors~\cite{Shor95,Steane96b,NielsenChuang}. All of the qubits are then replaced with encoded qubits and all operations are replaced with their encoded versions. This technique was looked at in \cite{Perseguers08b} for a square network using a majority voting CSS code. In this case for bonds containing $2t+1$ edges the phase error probability on each encoded edge can be suppressed to $\approx \binom{2t+1}{t+1} p_z^{t+1}$. To maintain a constant phase error probability on the final state the number of edges in each bond only needs to increase logarithmically with the size of the network. Unfortunately, this operation also increases the chance of a bit flip on the encoded edge to $\approx(2t+1)p_x$, which means there is a point where the chance of a bit-flip exceeds the network's critical value, patches are no longer small and the procedure fails. When $p_x$ is small this still enables the $p_z$ error probability to be substantially reduced. For rank two, binary states, the number of edges required in each bond does not need to scale with distance and yet for rank four states there is a logarithmic scaling. In the following we are interested in extending the constant scaling with states that are of rank three.

\section{Entanglement Distribution with Rank Three States \label{sec3}}
Using the global error correction procedure we can cause the edge
errors in every pure state in an ensemble to condense into patches on
a large GHZ state in such a way that a highly entangled two-qubit
state can be generated between two distant nodes. This requires the
initial network nodes to be linked by sufficiently entangled binary
states. However, in the following we will show that such states can be
probabilistically generated from a larger class of rank three states
using LOCC. This widens the class of states in a 2D network that
allow for long distance entanglement distribution.

\begin{figure}[]
  \centering\includegraphics[width=8cm]{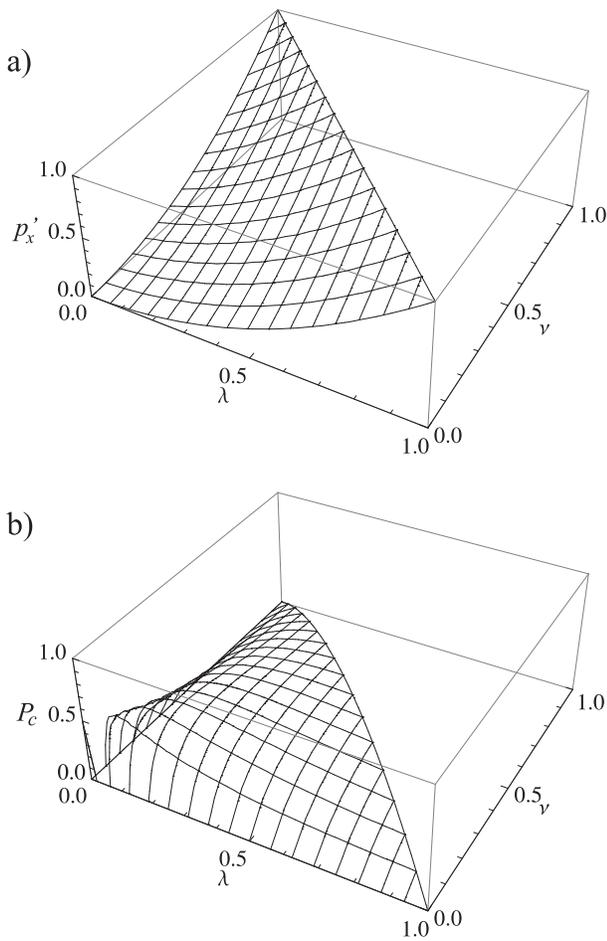}
\caption{%
a) $p_x'$ coefficient for the binary state created using a PCM. b) The probability, $P_c$, of generating a binary state from two initial states, i.e. $m=2$, of the type given by Eq.~\ref{eq9}.
   }
\label{fig7}
\end{figure}

It is impossible for a rank two, binary state to be
  created using anything other than states that are rank three or
  less. We give a short proof of this fact in appendix~\ref{app1}. If
we have two rank three states that can be probabilistically
transformed by LOCC into states of the form
\begin{equation}
\rho(\lambda,\nu) = \lambda \ket{\psi_{00}}\bra{\psi_{00}} + \nu \ket{\psi_{01}}\bra{\psi_{01}}+(1-\lambda-\nu)\ket{01}\bra{01},
\label{eq9}
\end{equation}
with $0\leq\lambda,\nu\leq 1$ and $\lambda+\nu\leq 1 $, then they can be probabilistically distilled into a binary state
\begin{equation}
\rho_b(p_x') = (1-p_x') \ket{\psi_{00}}\bra{\psi_{00}} + (p_x') \ket{\psi_{10}}\bra{\phi_{10}},
\label{eq10}
\end{equation}
where
\begin{equation}
p_x' = 1- \frac{(\lambda+\nu)^2+(\lambda-\nu)^2}{2(\lambda+\nu)^2}.
\label{eq11}
\end{equation}
The states given by Eq.~(\ref{eq9}) can actually form 
(up to local unitaries) when both qubits of a maximally entangled state, $\ket{\psi_{10}}$, pass through
phase flip and amplitude damping channels. Having two of these states
the transformation to a binary state is done by performing a protocol
discussed in Ref.~\cite{Broadfoot09,Broadfoot10}, where it was called
a `pure state conversion measurement' (PCM), followed by local Hadamard
operations. The PCM involves performing two C-NOT gates locally, with
one entangled state's qubits acting as the target qubits. These target
qubits are then measured in the computational basis. If we find both
qubits to be in the state $\ket{1}$ it has succeeded, which happens
with a probability of $(\lambda+\nu)^2/2$. If it fails then we destroy
the entanglement and lose both edges used. A graph of $p_x'$ and the
probability of succeeding is shown in Fig.~\ref{fig7}. From this it
can be clearly seen that to minimize the resulting $p_x'$ we would
like $\vert\lambda-\nu \vert$ to be maximal. The maximal success
probability with two states is $1/2$. If we have $m$ initial states in
a bond we can generate a binary state with finite probability
exceeding $P_c=1-(1-((\lambda+\nu)^2/2))^{\lfloor m/2 \rfloor}$, which
can be arbitrarily close to one given enough edges. Again, if this
fails then all of the edges are lost.

\begin{table}[]
\centering
\begin{tabular}{ll}
\hline
\hline
 Lattice & Threshold $p_c$ \\
 \hline
 2D Square & 0.5 \\
 2D Triangular & $2\sin(\pi/18) \approx 0.347$ \\
 2D Honeycomb & $1-2\sin(\pi/18) \approx 0.653$ \\
 \hline
 \hline
 \end{tabular}
\caption{Threshold probabilities for various regular network geometries~\cite{book,Lorenz98}.
\label{tab1}}
\end{table}

\begin{figure}[t]
  \centering\includegraphics[width=9cm]{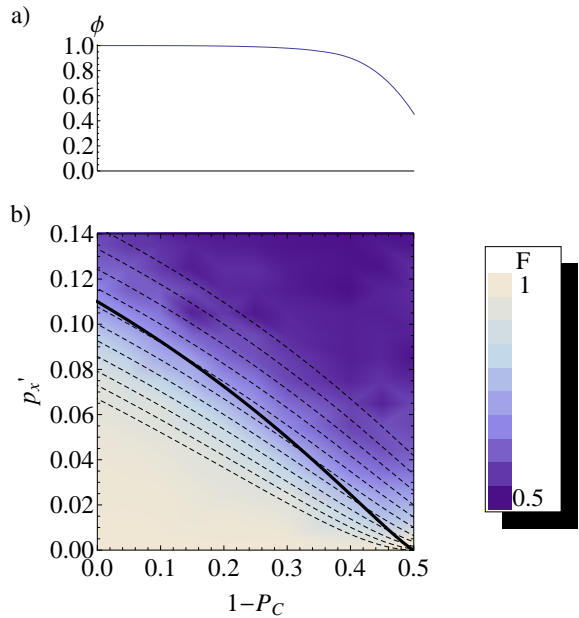}
\caption{%
  (color online) a) The probability of a node existing in the largest
  cluster, $\phi$. b) Based on a 25x25 square network, the fidelity
  $F$ between two randomly chosen nodes in the largest cluster
  of connected nodes is shown
  depending on the bond conversion probability
  $P_c$ and $X$-error probability $p_x'$. The boundary for values that
  allow long distance entanglement distribution in infinite networks
  is shown as a thick black line. Pure states can be
  probabilistically transformed into binary states with parameters
  along the dashed lines. These lines are for the pure states
  $\alpha=0.75, 0.76 \cdots 0.85$ from lowest to highest $p_x'$
  values, respectively.}
\label{fig8}
\end{figure}

By using this distillation procedure it is possible to transform an
initial network composed of rank three states into a network of binary
states, where the bonds are missing with a certain probability. This is
the case for the network shown in Fig.~\ref{fig5}. All of the nodes that are connected by a path of entangled states are then said to be in a cluster. For entanglement to be generated between two nodes they must both be in the same cluster and this links with results from bond percolation theory \cite{book}. For a square network of infinite extent, if the success probability of transforming each bond into a
binary state is larger than $1/2$ then a cluster that contains
infinite nodes is almost certain to exist. This is the `percolation
threshold' and values for further geometries are given in
Tab.~\ref{tab1}. The probability of a node being inside the infinite
cluster is given by the `percolation probability', $\phi(P_c)$. The
values of these together with $\psi(P_c)$, the probability that a bond
exists and is linking nodes in the infinite cluster, can be calculated
numerically for different
networks~\cite{Perc1-RevModPhys.45.574,Perc2-PhysRev.126.949}. For
large finite networks this threshold phenomenon still exists but the
transition becomes smoother. In this case, when the critical value is
exceeded we obtain a large cluster that contains a majority of the
nodes. On this cluster we can then apply our general global correction
procedure to obtain a highly entangled state between two nodes. This
requires that the binary states present in the largest cluster are
sufficiently entangled for global error correction to succeed. We use
entropic arguments to give an approximate threshold for a finite
square network, with sides consisting of $L$ nodes. The cluster will
contain an average of $\left\langle N_E\right\rangle=2L(L-1)
\psi(P_c)$ binary states, with each of these introducing a Shannon entropy $H(p_x')$. The measurements then extract one bit of information from every plaquette
in the cluster. The number of these plaquettes on average is given by
$\left\langle N_P\right\rangle=2+\left\langle
  N_E\right\rangle-\left\langle N_N\right\rangle$, with $\left\langle
  N_N\right\rangle=L^2 \phi(p_x)$ being the average number of nodes in
the cluster. We require $\left\langle
  N_P\right\rangle\,>\,\left\langle N_E\right\rangle H(p_x')$ for
enough information to be gathered and this gives a bound for $p_x'$
when we solve $H(p_x')=\left\langle N_P\right\rangle/\left\langle
  N_E\right\rangle$. These bounds define a region for $P_c$ and $p_x'$
within which long distance entanglement distribution can be achieved.
In the infinite network case, $N \rightarrow \infty$, we have
$\psi(P_c)=P_c \phi(P_c)$ and can calculate that
\begin{align}
\left\langle N_P\right\rangle/\left\langle N_E\right\rangle &=1+\frac{2-L^2\phi(P_c)}{2L(L-1)P_c \phi(P_c)} \nonumber \\
& \stackrel{N\rightarrow \infty}{\rightarrow} 1-(2P_c)^{-1}.
\label{eq12}
\end{align}
 This defines the region by the relation $(1-H(p_x'))P_c>1/2$ and the boundary is shown in Fig.~\ref{fig8}. Our procedure is also related to methods developed to cope with qubit loss in surface codes \cite{Stace09} and similar critical regions have been found in these situations. For the case where $P_c=1$ the critical value of $p_x'$ in an infinite square network is then given by $H(p_x')=1/2$. This can be solved to yield a threshold of $p_x'=0.11$. Further analysis for the complete square case is performed in Ref.\cite{Perseguers08b} and by relating the scheme to surface code error correction~\cite{Dennis02} a threshold of $p_x'=0.1094$ is found. Similarly, in a complete, infinite triangular network we have $H(p_x')=\left\langle N_P\right\rangle/\left\langle N_E\right\rangle=1+(2-L^2)/(L-1)(3L-1)\rightarrow 2/3$ yielding a threshold of $p_x'=0.17$. Both of these thresholds are shown on Fig.~\ref{fig6}. In this figure it can be seen that the higher connectivity present in the triangular network does provide for higher fidelity and a lower threshold than a square network. The simulations were run for finite networks so a perfect transition at the critical threshold does not occur. However, as the networks become larger the fidelity will approach $1/2$ for our critical values. Note that these critical values are calculated for the ideal method and will be slightly larger than the simulated threshold due to the random choice of paths.
 
 We have performed numerical simulations of the
 scheme, when edges are removed randomly, for a 25x25 square network.
 The final singlet state fidelity for different distillation
 parameters was calculated and the results are shown in
 Fig.~\ref{fig8}. Our boundary between the values allowing long distance entanglement distribution and those that do not is also shown. This gives a good separation between the parameters producing singlets and those resulting in separable states. There is no sharp transition between the low and high fidelity region which is due to the finite network size.
 The limiting cases of pure state percolation and
   binary state global error correction are given by $P_c>0.5$,
   $p_x'=0$ and $P_c=1$, $p_x'<0.11$, respectively.

\section{Tradeoffs between different strategies on Pure state networks  \label{sec4}}
It is interesting to look at the behavior of entanglement percolation
and global error correction when creating long distance entanglement
from the same initial network. All of these procedures
require a network of identical initial states. If
the bonds in the network contain enough entanglement we produce a highly entangled state between two nodes that are a long distance apart. The two nodes that are connected are out of a random
set of nodes. Different methods may have benefits such as requiring
less entanglement in the initial bonds, generating higher entangled
states between two nodes or allowing a greater proportion of nodes to
be linked by the entangled state. Pure states are the simplest case
where this comparison can be considered so we will look at this case.
By randomly generating states between nodes from the
  initial states and then using global error correction we have found
  a method that works on networks containing rank three states. By
  changing the states generated we can apply this scheme on networks
  of pure states, together with global error correction and
  entanglement percolation. We will see that there is a wide choice of
  states to generate and each of these may provide further benefits
  compared to entanglement percolation or global error correction.

For pure non-maximally entangled states entanglement percolation is
known to succeed in creating a long distance singlet state
deterministically~\cite{Acin07}. This requires a geometry dependent
amount of entanglement, related to the percolation thresholds (see
Tab.~\ref{tab1}). The pure states are characterized by their
\emph{Schmidt decomposition}
\begin{equation}
\ket{\alpha}=\sqrt{\alpha}\ket{00}+\sqrt{1-\alpha}\ket{11},
\label{eq13}
\end{equation}
with $\alpha$ being the largest Schmidt coefficient,
$1\geq\alpha\geq1/2$. Any pure state can be put into this form via
local unitary operations and $\alpha$ can be considered as a measure
of the state's entanglement.

\begin{figure}[]
  \centering\includegraphics[width=8cm]{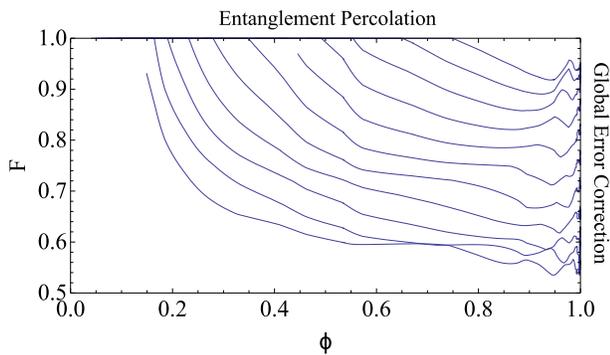}
\caption{%
  (color online) Starting from pure states we can create a range of
  binary states (dashed lines in Fig.~\ref{fig8}). For each of these
  the largest cluster contains a different proportion $\phi $ of total
  nodes and any two of these can be linked using global error
  correction to leave a two qubit entangled state that has fidelity
  $F$. The different values of $F$ and $\phi$ are given for each
  initial pure state. Shown are plots corresponding to eleven initial pure states with 
  $\alpha=0.75,\,0.76,\,0.77,\ldots,0.85$ (top to bottom).}
\label{fig9}
\end{figure}

Apart from basic entanglement percolation other techniques have also
been proposed in pure state networks with smaller initial entanglement
requirements \cite{Acin07,Perseguers08,Lapeyre09,Perseguers09b}. These
include the global error correction on a square network with no
missing edges, as discussed in \cite{Perseguers08b}, and it was shown
that a lower amount of entanglement was required to succeed at the
expense of the final states' fidelity. By randomly
  generating different states before global error correction it
  becomes possible to explore the transition between global error
  correction and entanglement percolation. Here we will again choose
  to create binary states by probabilistically converting pure states
  into the binary states.  When we start with pure states there are a
  number of ways to perform this operation. These create the binary
  states with higher probability of success or more
  entanglement. To perform the conversion each pure state is
transformed, using LOCC, into another pure state
  \begin{equation}
\ket{\alpha'}=\sqrt{\alpha'}\ket{00}+\sqrt{1-\alpha'}\ket{11},
\label{eq14}
\end{equation}
that has higher entanglement with the optimal probability of
succeeding $P_c=(1-\alpha) / (1-\alpha')$. This operation involves a
local measurement at one qubit. The result is then communicated to the
other qubit where a unitary is performed. This transformation
operation and probability come from `Majorization'
results~\cite{Nielsen01}. These states can then be
`twirled'~\cite{Bennett96a,Vollbrecht01} into binary states, with
$p_x'=(\sqrt{\alpha'}-\sqrt{1-\alpha'})^2/2$. These
  actions probabilistically generate binary states at random bonds,
  allowing global error correction to be applied. The value of $P_c$
can be adjusted between $0$ and $1$. If $P_c$ is larger than the
percolation threshold then an infinite cluster, of binary states
$\rho_b\left(p_x'\right)$, is formed and a node is a member of this
cluster with probability $\phi (P_c)$. Any two nodes in the cluster
can then be linked by global error correction. Two nodes are in the
cluster with a probability of $\phi(P_c)^2$ and they can be turned
into a binary state that has fidelity $F$ after global error
correction. In Fig.~\ref{fig8} the paths of
$\left(1-P_c,\,p_x'\right)$ are shown for different values of $\alpha$
as a dashed line. Each point on these lines correspond to a binary
state that can be obtained from the initial pure state. The
relationship between the available fidelities, $F$, and fraction of
nodes in the cluster is given in Fig.~\ref{fig9}. Each line relates to
a different initial pure state. For each dashed line
  representing an initial state in Fig.~\ref{fig8} the only points
  allowing long distance entanglement in the infinite case lie below
  the critical boundary, shown in Fig.~\ref{fig8}.
 
If we are required to generate long distance
  entangled states with the highest possible amount of entanglement
then, given that the threshold for entanglement percolation is satisfied,
the preferred method is to randomly obtain perfect
singlets and then use entanglement swapping. This generates a perfect singlet, however in doing so we
reduce the number of nodes that are available to be linked. When the
threshold is not satisfied we must probabilistically
  transform the initial states to binary states with the largest
$1-P_C$ that still lie within the critical region. By doing this one
highly entangled state is created over a long distance. Note that the
initial state of the edges needs to be known in order to decide on the state
to percolate. We are unable to choose the nodes before running the
protocol, but we still know which nodes are linked at the end. If we
are required to link two particular nodes, that have been chosen
before the protocol is run, we must maximize $\phi$ and this is
achieved by using global error correction. The choice between
maximizing the fidelity or $\phi$ is dependent on the intended use for
the entangled state. If one entangled state between any two nodes in a
set is sufficient it may be beneficial to maximize $\phi$. In order to
judge the schemes a figure of merit can be constructed from $\phi$ and
the fidelity.  This can then be maximized to reveal the preferred
scheme. For example a suitable figure of merit in the case where two
specific nodes need to be entangled would be $F \phi^2$ and this is
maximized for global error correction.

\section{Conclusion \label{sec5}}
 Previously, entangled states could only be generated over an arbitrarily long distance from rank two states, or less, in regular 2D networks\cite{Acin07,Perseguers08b,Broadfoot09,Broadfoot10}. Here we have devised a global error correction scheme that enables a highly entangled state to be generated over an arbitrary distance. This procedure can be applied to any network composed of binary states and allows the creation of multi-qubit GHZ states. By combining this with entanglement percolation we have extended this ability to any network and a class of rank three states. These states would result from a combined amplitude damping and phase error channel. For networks composed of these rank three states it becomes possible to produce highly entangled states over arbitrary distances, with constant resources between the nodes. Although this is still a restricted case it is one step closer towards 2D networks composed of full rank states and allows us to deal with two important kinds of noise. Initially there must be a sufficient amount of entanglement between the nodes for the scheme to succeed and we have provided an entropic estimate for the requirements in percolated networks, based on the estimate in Ref.~ \cite{Perseguers08b}. We have numerically calculated the resulting fidelities and the results agree with the threshold estimate. The behavior of both entanglement percolation and global error correction was previously investigated separately but here we have provided a method to connect these possibilities and in doing so revealed the preferred methods for different priorities. It is still an interesting question to ask whether long distance entanglement distribution can exist in a 2D network of Werner states even when the gates are assumed to be perfect.

 This research was supported by the ESF via EuroQUAM (EPSRC grant EP/E041612/1).

\appendix

\section{Impossibility to generate binary states from rank four states \label{app1}}

Binary states can not be generated from two-qubit full rank states and we prove this by contradiction. If there were to exist some local procedure that produced some binary states from a full rank state then, by acting after with local projective measurements on all qubits except those in one binary state, we can easily construct a method to generate a single binary state and pure separable state. This local procedure can be described by an operator $M_A \otimes M_B$, where $M_A$ and $M_B$ act locally. The dimension of the entire Hilbert space is $D$. They must also satisfy $M_A M_A^\dagger$ and $N_B N_B^\dagger \leq 1 $. The initial mixed state is
\[\sum_{i=1}^D p_i \ket{i}\bra{i},\]
with $ p_i>0, \sum_i p_i$ = 1 and $\ket{i}\in H_A \otimes H_B $ is a complete basis. The procedure can then convert this initial state into
\[\sum_{i=1}^D p_i M_A \otimes N_B \ket{i}\bra{i} M_A^\dagger \otimes N_B^\dagger = p \rho_{bin} \otimes \rho_{sep}.\]
$\rho_{sep}$ is a pure separable state, $\rho_{bin}$ is the two-qubit binary state and $p$ a non-zero probability. Since the initial state was full rank this property requires $M_A\otimes M_B$ to be rank two. This is only the case when either $M_A$ or $M_B$ have rank one. Any operator, $M_A\otimes M_B$, with this property would not generate an entangled state. Hence, a binary state can not be generated using LOCC from a full rank state and this includes the case of finite full rank two-qubit states.

\end{document}